\documentclass{PoS}
\usepackage{latexsym}
\usepackage{amssymb}
\usepackage{epsfig}
\usepackage{amsfonts}
\usepackage{amsmath}
\newcommand{\beqa}{\begin{eqnarray}}
\newcommand{\eeqa}{\end{eqnarray}}
\def\be{\begin{equation}}
\def\ee{\end{equation}}
\newcommand{\eqn}[1]{(\ref{#1})}
\newcommand{\del}{\partial}
\newcommand{\tr}{{\rm Tr}}
\newcommand{\jj}{{\mathbb P}}
\newcommand{\rd}{\mathrm{d}}
\newcommand{\TT}{{\mathbb T^{\gamma}}}
\newcommand{\cG}{{\mathcal G}}
\title{A field-theoretic approach to Spin Foam models in Quantum Gravity}

\ShortTitle{GFT models}

\author{\speaker{Patrizia Vitale}\\
        Dipartimento di Scienze Fisiche, Universit\`a di Napoli Federico II\\
and INFN, Sezione di Napoli, Via Cintia 80126 Napoli, Italy\\
        E-mail: \email{vitale@na.infn.it}}

\abstract{We present an  introduction to Group Field Theory models, motivating them on
the basis of their relationship with discretized BF models of gravity. We derive the Feynmann rules and compute  quantum corrections in the coherent states basis.}

\FullConference{Corfu Summer Institute on Elementary Particles and Physics - Workshop on Non             Commutative Field Theory and Gravity,\\
        September 8-12, 2010\\        Corfu Greece}

\begin{document}

\section{Introduction}
In this article we review, in a very elementary way, some of the  motivations which are at the
basis of a class of quantum field theories which have been proposed to describe quantum
gravity \cite{boul, ooguri, Freidel, oriti}.  The base manifold where these  models are defined
is represented by $D$ copies of the Lorentz group, where $D$ is the physical space-time
dimension, and the Lorentz group is replaced by rotations when the signature of the
metric is chosen to be Riemannian. For this reason they are known as Group Field
Theories (GFT). They can also be viewed as higher rank tensor field theories, thus
generalising matrix models which describe gravity in two dimensions, and sharing some
features with noncommutative quantum field theories. The main motivation for us to
investigate such models is that they provide a background independent formulation of
quantum gravity in which one sums both over topologies and geometries. Indeed, each
Feynman graph of a $D$ dimensional GFT can be dually associated with a specific
space-time triangulation. Moreover, it can be seen to reproduce exactly spin-foam
amplitudes \cite{baez}, although in a different basis.

The simplest group field theories correspond to quantization of the BF models, hence to
topological versions of gravity. The first non-topological model is the Barrett-Crane
model \cite{BC}. Recently new spin foam models have been proposed for the quantization
of the full 3+1 dimensional gravity \cite{FreKra,ELPR}. These models stem from an
improved analysis of the Plebanski simplicity constraints. These new theories could be
called dynamical since their propagators, combining two non commuting projectors, have
non-trivial spectrum. In this paper, after an introductory part dedicated to motivations
and to the discretization of gravity along the lines of lattice gauge theories, we focus
on such new models, derive the Feynmann rules and outline the computation of quantum
corrections in the coherent states basis. A detailed account of the results may be found
in \cite{noi}.

The paper is organized as follows. In  section 2 we review the discrete path integral derivation of 2+1 gravity in
the BF description. In section 3 we introduce the main tools of GFT, which we apply to
$SO(D)$ BF theories is section 4. We compute Feynmann amplitudes of such models and we
find that they coincide, for $D=3$, with the partition function of 2+1 gravity that we
have found in section 2. This serves as a motivation for the description of quantum
gravity in terms of GFT models. Therefore we review in section 5 the first order
formalism for the gravity action in 3+1 dimensions in the Holst formulation and we
study, in section 6, a GFT model for such an action. Finally, in section 7, we describe the stationary phase method used in
\cite{noi} to evaluate the asymptotic large spin regime of certain graphs.

\section{\bf {Discrete Gravity in 2+1 dimensions}}
In this section we briefly review the first order formalism for gravity in 2 + 1
dimensions and its discretization, to arrive at a discretized path integral. The
partition function so obtained will be the basis to introduce and motivate  GFT models, which reproduce the same result.

In 2+1 dimensions the first order action of gravity  has the form of a BF action with
gauge group $G= SO(2,1)$ ($SO(3)$ if the metric is Riemannian)
\be
S= \int \tr (e\wedge F[\omega])=\int \epsilon_{IJK} e^I F^{JK}[\omega]
\ee
where $e$ (the B field) is a vector valued one form on the
space-time manifold $M$ and $F$ the curvature of the   connection
one form, $\omega$, valued in the Lie algebra
$\mathcal{L}=so(2,1)$ ($so(3)$ in the Riemannian case). In the
language of fiber bundles $\omega$ is the connection one form of
the principal G-bundle,
 while $e$ is a one form on the
associated vector bundle, it is therefore vector valued. Once we
have chosen a trivialization, we can identify the fibers of the
vector bundle with the Lie algebra, so that $e$, $\omega$ and $F$
are all $\mathcal{L}$ valued
\beqa
e&=&e_\mu^I dx^\mu \tau_I \\
\omega&=& \omega_\mu^I dx^\mu \tau_I\\
F&=&F_{\mu\nu}^I dx^\mu\wedge dx^\nu \tau_I
\eeqa
$\tau_I\in \mathcal{L}$.
 Therefore, $\tr(e\wedge F)$ is the
3-form constructed by taking the wedge product of the differential form parts of e and F
and using the Killing form $<.,.>=-\tr(..)$  to pair their Lie algebra valued parts. The
 trace is taken in the adjoint representation.

In order to discretize this action on a lattice, let us consider a triangulation, $T$,
of the space-time manifold in terms of tetraedra, with faces  $f$, edges $\ell$ and
vertices $v$. We associate to the one form $e$ the Lie algebra element
\be
 E_\ell=\int_\ell e
\ee
To get the discrete analogue of the curvature as a quantity which is still associated to
the edges we need to introduce a dual lattice, $T_*$.  This is defined on associating to
each tetraedron a vertex in the middle, $v_*$, to each face an edge which intersects it,
$\ell_*$, to each edge a face $f_*$ whose boundary $\del f_*$ encircles the edge, to
each vertex a closed surface
 which encloses it. The two-dimensional  subcomplex
contained in such a cellular complex, whose building blocks are vertices, edges and faces, draws
a graph which we indicate with $\mathcal{G}$, also known in the literature as dual
skeleton.
 Therefore we associate to the
connection $\omega$ the group element
  \be
  h_{\ell_*}= P\exp\int_{\ell_*} \omega
  \ee
  and the discretized curvature  is defined as the holonomy of the connection along the boundary of $f_*$. Dual faces are
  associated to  edges of the direct triangulation, therefore we have:
\be
H_{f_*} \equiv H_\ell= \prod_{l_*\in \del f_*(\ell)} h_{\ell_*} \label{holon}
\ee
The discretized BF action becomes then
\be
S(E_\ell, H_\ell)=\sum_{\ell\in T} \tr E_\ell H_\ell
\ee
where the trace is to be taken in the fundamental representation of the group. We can define a discretized
partition function
\be
Z[T,T_*] =  \int_{\mathcal {L}}\prod_{\ell\in T} d E_\ell\int_G
\prod_{\ell_*\in T_*} dh_{l_*} \exp(i\tr(E_\ell \prod_{\ell_*\in
\del f_*}h_{\ell_*})
\ee
where the Lie group $G$ is either $SO(2,1)$ or $SO(3)$, depending
on the signature of the space-time manifold, while $\mathcal{L}$
is the appropriate Lie algebra. The integral in the Lie algebra
can be performed and we get
\be
Z[T_*]=\int\prod_{\ell_*\in T_*} dh_{\ell_*} \prod_{f_*\in T_*}\delta(\prod_{\ell_*\in
\del f_*}h_{\ell_*}) \label{deltabf}
\ee
We will derive this result from a GFT and this will serve as a motivation to investigate GFT models as candidates for quantum gravity.
Before going into that, let us recall  that \eqn{deltabf} is also the starting point to
derive the correspondence of discretized gravity in the first order formalism  with spin
foams. All what is needed is the  decomposition of the delta function on the
group in terms of irreducible representations, labelled by the half-integer $j$
\be
\delta(h)=\sum_j (2j+1) \tr_j R^j(h)
\ee
where $\tr_j$ is a shorthand for $\tr_{V_j}$, the trace in the finite dimensional vector space $V_j$, base of the representation.
 Replacing into \eqn{deltabf} and decomposing the representation of the product of group elements into a
product of representations we finally get
\be
Z[T_*]= \prod_{f_*}\sum_{j_{f*}} (2j_{f*}+1) \prod_{v_*}\{6j\}=A_\mathcal{G}
\label{spinfoam}
\ee
where all group integrations have been performed. $\{6j\}$ are the Racah-Wigner 6j symbols. See for example \cite{boul} for normalization conventions. As indicated, this is exactly the
spin-foam amplitude $A_\mathcal{G}$ of the (dual) graph $\mathcal{G}$. It represents
then the transition from a given space geometry to another, as a sum over histories
connecting the two.

\section{Group field theory}
In this section we  introduce the basic tools of Group Field Theory (GFT) in any
dimensions. GFT has been first introduced in \cite{boul, ooguri}, then further  studied by
\cite{Freidel,  oriti},
 to describe discretized models of Quantum Gravity in terms of fields living on the dual lattice of the discretization.
 We first introduce the main ingredients of the theory, such as the notion of field on the group manifold,
 the propagator and the interaction and then we motivate our definitions showing that the quantum field theory so defined
generates, in a different basis,  the spin foam amplitudes of gravity in 2+1 dimensions.

In our approach all the dynamical content of the models considered is encoded in the
propagator, while the interaction vertex is fixed by the request that it be the simplest
generalization of local vertices of quantum field theory.

\begin{figure}
\centerline{\epsfig{figure=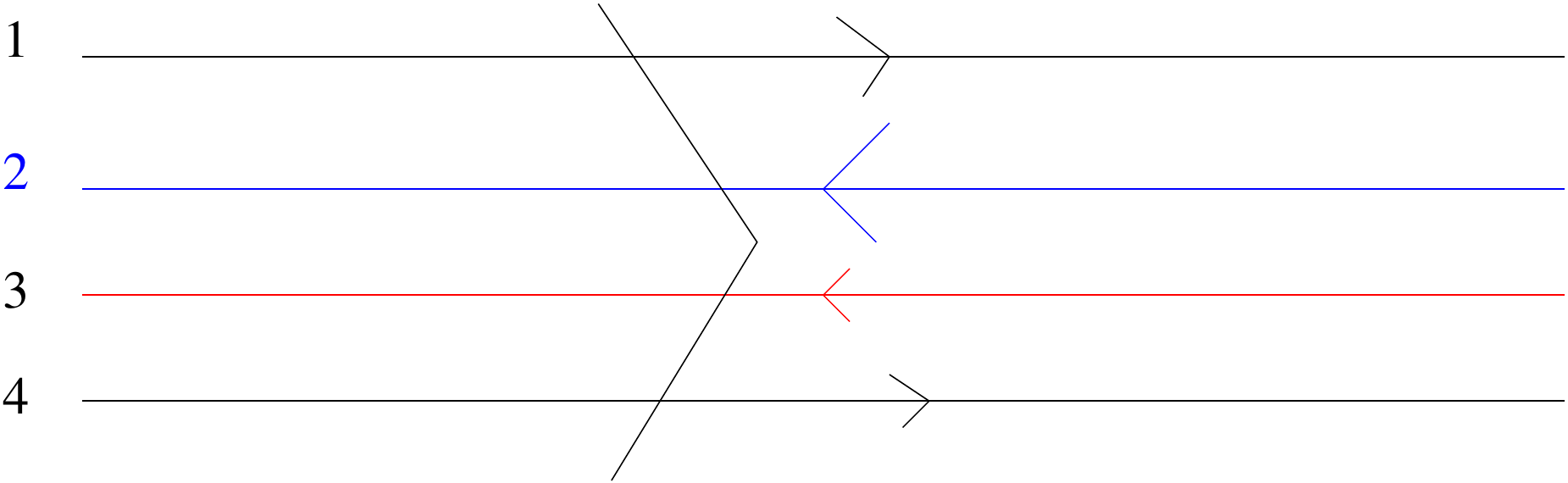,width=4cm, height=1cm} } \caption{A stranded
propagator in D=4 with particular orientation; two strands have $\eta_{\ell f}=1$ and
the other two have $\eta_{\ell f}=-1$.} \label{propa-fig}
\end{figure}

\begin{figure}
\centerline{\epsfig{figure=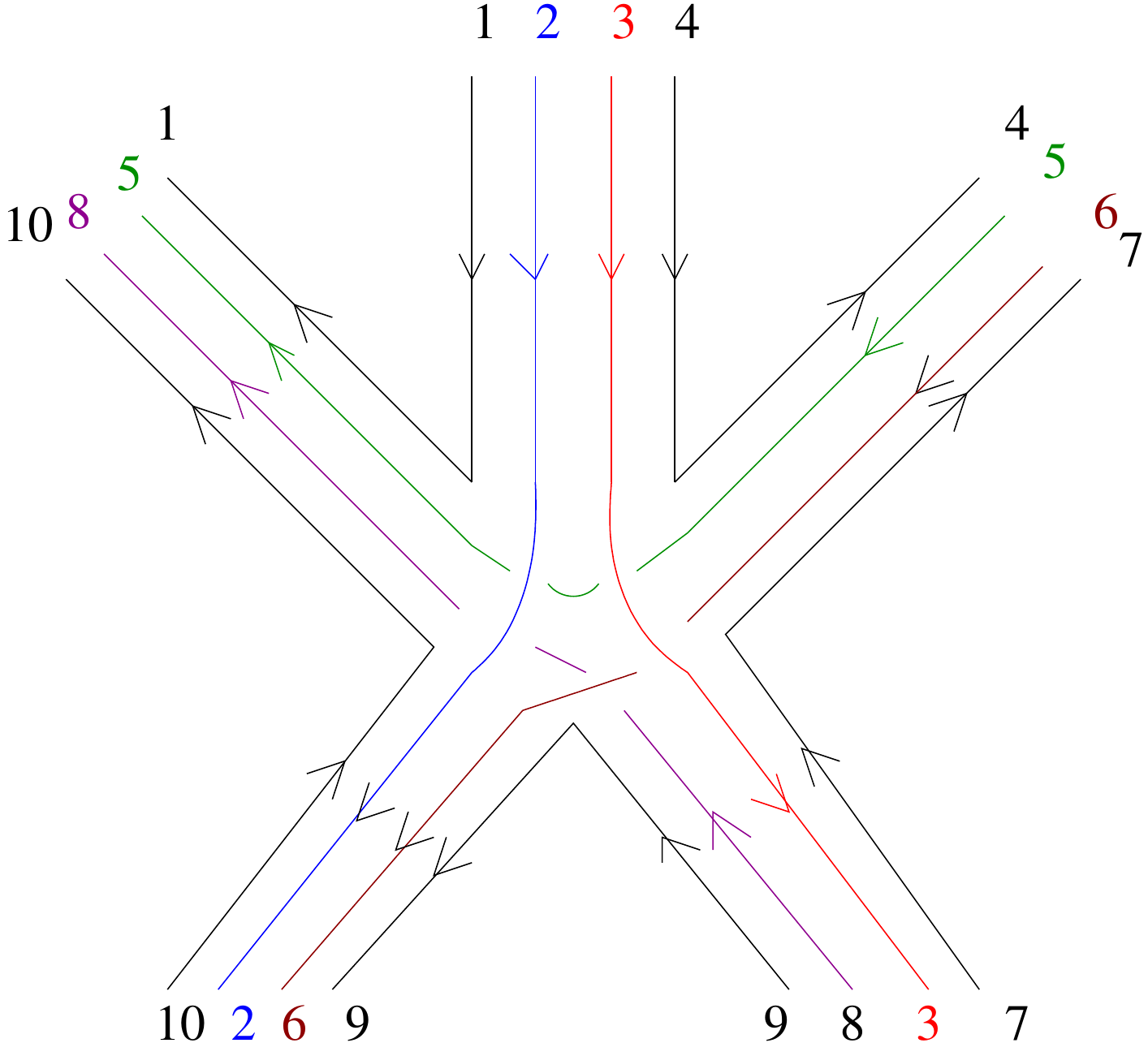,,width=4cm,height=3cm}  } \caption{A simple
vertex of a $4-$dimensional GFT. We have chosen here a particular matching and
orientation for each of the strands.} \label{vertexf}
\end{figure}

\begin{figure}
\centerline{\epsfig{figure=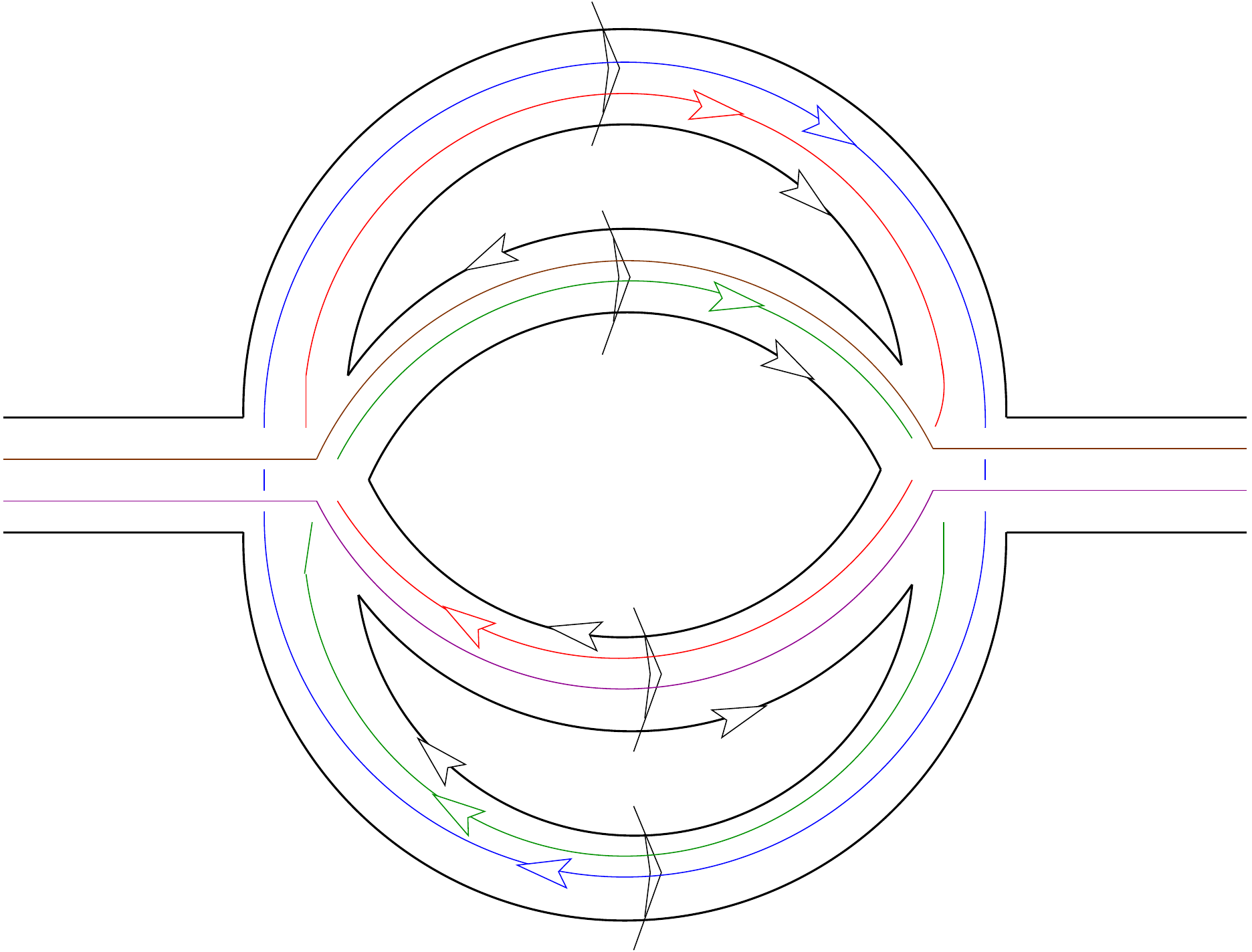,width=5cm, height=3cm} } \caption{The ``self-energy''
graph $\cG_2$ in 4 dimensions, quantum correction to the propagator.} \label{2P-fig}
\end{figure}

  We work with Riemannian space-times, therefore the relevant Lie group will be
$SO(D)$. In three dimensions we will work with its double covering $SU(2)$. In four
dimensions we will use the splitting $SO(4)\simeq SU(2)\times SU(2)$. In GFT the field
arguments live on products of Lie groups
 \be
 \phi:  (g_1,...g_D)\in [SO(D)]^D\rightarrow \phi(g_1,...g_D)
 \ee
with $D$ the  spacetime dimension. The propagator is an Hermitean map
 $
 C:\phi\to C \phi  \;\;
 $
 with Hermitian
kernel $  \;C(g_1, \ldots, g_D; g'_1, \ldots, g'_D)$:
\be
 [ C\phi ] (g_1, \ldots , g_D) = \int  d g'_1  \ldots d g'_D C(g_1, \ldots, g_D; g'_1,
\ldots, g'_D) \phi (g'_1, \ldots , g'_D).
\ee
It is therefore represented by a stranded line with $D$ strands (fig. \ref{propa-fig}).
The precise form of $C$ is what distinguishes the different models. It encodes the
dynamical content of the theory. Vertices instead only depend on space-time dimensions
and are the same for all models. We define a {\it simple vertex} joining $2p$ strands in
terms of its kernel: this    is a product of $p$ delta functions matching strand
arguments, so that each delta function joins two strands in two different lines.
 The usual vertex for $D-$dimensional GFT is a $\phi^{D+1}$
simple vertex (see fig \ref{vertexf}). For instance the  $SU(2)$ BF vertex in 3
dimensions  is (with $p=6$)
 \be
S_{{\mathrm{int}}}[\phi]= \frac{\lambda}{4}\int \left(\prod_{i=1}^{12}d
g_i\right)\phi(g_1,g_2,g_3) \phi(g_4,g_5,g_6)\phi(g_7,g_8,g_9)
\phi(g_{10},g_{11},g_{12})\;
 K(g_1, .., g_{12}), \nonumber
\ee
with
\be  K(g_1, .. g_{12}) = \delta(g_3g_4^{-1})\delta(g_2g_8^{-1})\delta(g_6g_7^{-1})\delta(g_9g_{10}^{-1})
\delta(g_5g_{11}^{-1})\delta(g_1g_{12}^{-1})\label{vertex}
\ee
 A stranded graph (see fig. \ref{2P-fig}) is called {\it regular} if it has no {\it
tadpoles} (hence any line $\ell$ joins two distinct vertices) and no {\it tadfaces}
(hence each face $f$ goes at most once through any line of the graph). Although final
results do not depend on them, it is convenient to introduce orientations in terms of
incidence matrices:
\begin{itemize}
\item the ordinary incidence matrix $\epsilon_{\ell v}$ which has value $1$  $(-1)$ if the edge $\ell$ enters (exits)
 the vertex $v$, 0 otherwise.
\item the incidence matrix $\eta_{f \ell}$  which has value $+1$ if the face $f$
goes through the edge $\ell$ in the same direction, $-1$ if the
face $f$ goes through the edge $\ell$ in the opposite direction, 0
otherwise.
\end{itemize}
\section{GFT for $SO(D)$ BF theory}
In three dimensions gravity
is well described by a BF theory, with gauge group $SU(2)$. The discrete partition
function has been obtained in \eqn{deltabf}. Such a result is indeed valid for BF in any
dimension.  We wish to reproduce such a result with a GFT model. In any dimension the
propagator in direct space (that is in group variables) is just the projection on gauge
invariant fields,
\be
\jj (\phi)=\int_{SO(D)} d h \phi(g_1 h,\ldots, g_D h),
\ee
 $\jj ^2=\jj$ so that the
only eigenvalues are $0$ and $1$ (which means that the BF theory has
no dynamics).
 $\jj$ is Hermitian with kernel
\be \jj (g_1, ..., g_D ; g'_1, ... g'_D)=\int d h \prod_{i=1}^D
\delta (g_i h (g'_i)^{-1}). \label{bfprop}
\ee
Let us look at  amplitudes in direct space. We choose an arbitrary
orientation of the lines and faces of a graph $\mathcal G$ (which
for simplicity has no external legs). Combining the vertex  and
the propagator, integration over all strand variables $g, g'$ leads to
\be
A_{\mathcal G} = \int \prod_{\ell \in {L}_{{\mathcal G}} } d
h_\ell \prod_{ f \in {F}_{{\mathcal G}}} \delta \left(
{\vec {\prod}_{\ell \in f} h_\ell^{\eta_{\ell f} }}   \right). \label{amp}
\ee
where ${L}_{{\mathcal G}}$, ${F}_{{\mathcal G}}$ are the set of lines and faces of
${\mathcal G}$, respectively. The oriented product $\vec{\prod}_{l \in  f} h^{\eta_{\ell
f}}$ means that the product of the variables $h_\ell$ has to be taken in the cyclic
ordering corresponding to the face orientation (starting anywhere on the cycle). As
announced, the amplitude (\ref{amp}) neither depends on the arbitrary orientation of the
lines, nor on those of the faces. This result is valid for any $D$. In order to compare
it with our previous results in discrete gravity we have to understand the graph
$\mathcal{G}$ as a graph on the dual discretization of space-time. Then
 we can check that this is
exactly the result we have found for the partition function of discretized BF
\eqn{deltabf}.\footnote{ We will understand all GFT graphs as living in the dual
triangluation from now on.} Therefore the Feynmann amplitudes of GFT with propagator and
vertex given respectively by \eqn{bfprop} and \eqn{vertex} reproduce exactly the
partition function of BF models, which also means, in a different basis, spin foam
transition amplitudes.

 As we will see in the next section, in 3+1 dimensions gravity is
described by a constrained BF theory. The difficulty in implementing the constraints at
the discrete level  translates into an analogous difficulty in proposing an unambiguous
GFT model.

\section{Gravity in 3+1 dimensions: the Palatini-Holst action}
It is very well known that, as for 2+1 dimensions, we can describe General Relativity with a first order
formalism which is, at least classically, equivalent to the second order formalism of
the Einstein-Hilbert action. Differently from the 2+1-D case, it is not just a BF
theory, but we have to implement constraints. The first order action reads
\be
S=\frac{1}{8\pi G}\int_M \epsilon_{ABCD} e^A\wedge e^B \wedge F^{CD}[\omega]
\ee
where $\omega=\omega_\mu^{AB} dx^\mu \tau_{AB}$ is the  connection one-form, valued in
the Lie algebra of $SO(3,1)$, with generators    $\tau_{AB},\; A=1,..4$ ($4\times 4$
antisymmetric matrices). $F=F^{AB}_{\mu\nu}dx^\mu \wedge dx^\nu\tau_{AB}$ is the
curvature two-form; it is Lie algebra valued. The  tetrads $e=e^A_\mu  d x^\mu v_A$ are
vector valued one-forms such that their wedge product may be regarded as $so(3,1)$
valued, once a trivialisation chosen
$
e\wedge e= e^A_\mu e^B_\nu  dx^\mu\wedge dx^\nu \tau_{AB}.
$
Therefore, similarly to the 2+1 dimensional case, the action is entirely described in
terms of Lie algebra valued forms, the curvature $F$ and the two form $e\wedge e$, whose
wedge  product is a 4-form which is integrated on the manifold $M$; as for their Lie
algebra dependence, on identifying respectively the generators of rotations and boosts
with $J_a, P_a, a=1,..,3$  we can use the nondegenerate bilinear form on  the algebra, $<.,>_1$
\be
<J_a,P_b>_1=\delta_{ab}, \;\; <J_a,J_b>_1=<P_a,P_b>_1=0
\ee
 to rewrite the action as
\be
S_1=\frac{1}{8\pi G}\int_M \tr_1 (e\wedge e \wedge R)\label{action1}
\ee
with $\tr_1:= <.,.>_1$. It can be easily checked that it generates the correct equations
of motion.
The peculiar characteristic of the group $SO(3,1)$ and of its Euclidean version $SO(4)$
is that there is another invariant quadratic form on the Lie algebra, $\tr_2=<.,.>_2$,
so defined
\be
<J_a,J_b>_2=<P_a,P_b>_2=\delta_{ab}, \;\; <J_a,P_b>_2=0\label{metric2}
\ee
(actually, eq. \eqn{metric2} is the specialization to 3+1 dimensions of a quadratic form
that would exist for any d+1).  This allows to pair the  two-forms $F$ and $e\wedge e$
in a different way so that a new action is produced
\be
S_2=\frac{1}{8\pi G} \int_M \tr_2(e\wedge e\wedge F)=\int_M e^A\wedge e^B \wedge
F_{AB}[\omega] \label{action2}
\ee
with $\tr_2:=<.,.>_2$. It can be easily shown that this new term can be safely added to
the gravity action \eqn{action1} without changing the equations of motion. Although we
have no classical consequences, we have to expect a different quantum theory. The full
action $ S_H= S_1+ \frac{1}{\gamma}S_2 $ is known as the Palatini-Holst action
\cite{holst}, while the real parameter $\gamma$ is the Barbero-Immirzi parameter
\cite{barbero, immirzi}. $S_H$ can be regarded as a constrained BF action
\be
S_H=\int_M B^{AB}\wedge F_{AB} + \phi(B) \label{sh}
\ee
where the constraint has to implement
\be
B^{AB}=\epsilon^{AB}_{CD}e^C\wedge e^D+\frac{1}{\gamma} e^A\wedge e^B. \label{constraint}
\ee
In the so called Plebanski formulation the constraint is of the form $ \phi(B) \propto
\phi_{ABCD}B^{AB} \wedge B^{CD}. $ The action is therefore dependent on three variables,
$B,\omega,\phi$. Variations w.r.t. to all of them essentially give back  Einstein
equation.

The discretization of the 4D BF action $S=\int_M B^{AB}\wedge F_{AB}$   goes along the same lines as the 3D case. We have
a triangulation T of the space-time manifold with a 4-D simplicial complex, and a dual
lattice $T^*$ with its 2-dimensional subcomplex which identifies our graphs. The
$B$ field now being a 2-form, has to be integrated over faces $f$ to produce $so(4)$
(Lie algebra) elements, $B_f$. The discretized curvature  is defined as in 3D to be the
holonomy of the connection along the boundary of $f_*$ \eqn{holon}. Dual faces are
 now  associated to  faces of the direct triangulation, therefore we have:
\be
S(B_f, H_f)=\sum_{f\in T} \tr B_f H_f
\ee
When writing the partition function for such a discretized action, upon integrating over the Lie algebra, we end up with the
same result as in \eqn{deltabf} with the gauge group now $SU(2)\times SU(2)$. When using
the Peter Weyl theorem to decompose the group delta function on irreducible representations, we obtain a spin
foam amplitude which is the analogue of \eqn{spinfoam}, with 6j symbols now replaced by
15j symbols (see \cite{ooguri} for normalization conventions) and $j\equiv(j_+,j_-)$, according to the splitting of $SO(4)$.

To treat the constrained action \eqn{sh} we have to include the
Plebanski constraint.  There are various proposals on how to implement it at the discrete
level. We have considered just one of them, the EPRL spin-foam model \cite{ELPR}, in the
 Freidel-Krasnov formulation \cite{FreKra}, which uses coherent states. This is the argument of the next section.

 \section{GFT approach to the EPRL-FK model}
 In our approach all the dynamical content of different models is encoded in the propagator, while the vertex
 remains the same as in BF theory \eqn{vertex}. Therefore the constraint is directly imposed on the propagator
 \eqn{bfprop}. This is better understood in the coherent states basis.
In 4 dimensions  we use the splitting of $SO(4)$ as the product $SU(2)\times SU(2)$
 and we decompose the group elements as $g= (g_+ , g_-)$, with $
g_{\pm}\in SU(2)$.  In 3D we use $SU(2)$ coherent states
$
|j,g > \equiv g |j,j> = \sum_m  |j,m > [R^{(j)}]^m_{j}(g) .
$
with $R^{(j)}(g)$ the 2j+1 dimensional irreducible representation of g. The identity is
decomposed over coherent states as
\be
1_j = \rd_j \int_{{\rm SU}(2)} dg \, |j,g >< j,g|= \rd_j \int_{G/H = S^2} dn \, |j,n ><
j,n| \label{PU3}
\ee
with $|j,n>=R^{(j)}(g_n)|j,j>$.  In 4 dimensions  we indicate with ${j}\equiv(j_+,j_-)$  the eigenvalues of the
Lie algebra generators $J$ in each $SU(2)$ component and  we work with $SU(2)\times SU(2)$
coherent states
 $|j_+,n_+  > \otimes |j_-,n_- >$.  The identity is therefore
\be
1_{j}  =  \rd_{j_+} \rd_{j_-}\int_{S^2\times S^2} d n_+d n_- |j_+,n_+>\otimes |j_-,n_-> <
j_+,n_+|\otimes < j_-,n_-| \label{PU4}
\ee
As an intermediate step towards the full 4-dimensional case we rewrite the D-dimensional
BF propagator \eqn{bfprop} in a different way (D=3, respectively D=4).
 Since $\jj^2 = \jj$ we  introduce two
gauge-averaging  variables, $u, v \in G$ at the ends of the propagator,  ($u$ on the
side where $\epsilon_{v \ell}=-1$ and $v$ on the side where $\epsilon_{v \ell}=+1$). G
is either $SU(2)$ or $SU(2)\times SU(2)$ depending on the dimension. Between these two
variables we insert the partition of unity \eqn{PU3} (resp. \eqn{PU4})
\be
\jj (g ; g') =\int d u d v \prod_{f=1}^4 \sum_{j_{f}} \rd_{j_{f}}
 \tr_{V_{j_{f}}} \left( u g_f (g'_f)^{-1} v^{-1}
1_{ j_f}\right) \label{genprop}
\ee
where, to simplify the notation,  we omit to indicate the representation explicitly and use $g\equiv R^{(j)}(g)$ for all group elements, from now on.
This is equivalent to \eqn{bfprop} upon redefining $h= v^{-1}u$.
 The index $f$ labels the strands of the propagator according to the face to which they belong. In other words, on introducing
\be
\mathbb{I}=\oplus_{j_{f}}    \otimes_{f=1}^4  1_{j_f}
\ee
we have used
\be
\jj=\jj\mathbb{I} \jj \label{fancyprop}
\ee
This redundancy will be useful in a while to better understand the constrained propagator.

 The amplitude of a given graph ${\mathcal G}$ with a certain number of vertices and propagators may be written as a factorization over faces
 \be
A_{\mathcal G} = \int \prod_{\ell \in {L}_{{\mathcal G}} }d u_\ell d v_{\ell} \prod_{  f
\in \mathcal{F}_{{\mathcal G}}} {\mathcal A}_f \label{amp8}
 \ee
We assume for simplicity  that all faces are closed. Then, for each given face $f$ of  length $p$,   we   number the vertices and lines in the
(anti)-cyclic order along the face   as $ \ell_1, v_1 \cdots \ell_p, v_p$,
with  $\ell_{p+1} = \ell_1$. Combining vertices (delta functions for the strand group variables, $g_i, g'_i$) and propagators, and integrating over the strand group variables, $g_i, g'_i$ we obtain
\beqa
\mathcal{A}_{f} =\sum_j d_j^{p+1}\prod_{a=1}^p\tr_j  t_{\ell_a,  v_a  }^{\eta_{\ell_a
f}} t_{\ell_{a +1, v_a}}^{\eta_{\ell_{a+1}  f}} 1_j= \sum_j d_j^{p+1} \int\prod_{a=1}^p
d n_{\ell_a f} <j, n_{\ell_a f} |t_{\ell_a,  v_a  }^{\eta_{\ell_a f}} t_{\ell_{a +1,
v_a}}^{\eta_{\ell_{a+1}  f}}|j, n_{\ell{a +1} , f}> \label{amp9}
\eeqa
where $t_{\ell_a , v_a}$ is $v_{\ell_a}$ if $\epsilon_{\ell_a  v_a} =+ 1$ and
$u_{\ell_a}$ if $\epsilon_{\ell_a  v_a} = - 1$. (Remember that in D=4 $ \rd_j\equiv\rd_{j_{f+}}\rd_{j_{f-}}$).

\subsection{The EPRL/FK GFT}
 The EPRL/FK propagator has a structure similar to
\eqref{genprop} but with replacement of $\mathbb{I}$ by a non-trivial projector which
does not commute with $\jj$. This implies that it is not possible to recombine $u$ and
$v$ in a single gauge averaging variable $h$. It implements in two steps the Plebanski
constraints with a non-trivial value of the Immirzi parameter $\gamma$. We only consider
$0 < \gamma \le 1$ where the EPRL and  FK models coincide. Starting from the
 expression (\ref{genprop}) of the $BF$ propagator in the coherent states representation,
the first step adds the constraint $j_+/j_- = (1+\gamma) /(1- \gamma) $, $n_+=n_-$.
 The second step replaces in each strand of (\ref{genprop}) the identity
${\mathbf 1}_j$ by a projector $T_j^{\gamma}$ whose definition is
\be
T^{\gamma}_{ j}=   d_{j_+ + j_-} \bigl[  \delta_{j_{f-}/j_{f+} = (1-\gamma) /(1+ \gamma) }  \bigr] \int dn  |j_+,n >\otimes
|j_-,n > < j_+,n | \otimes < j_-, n | \label{TBFEPRL} .
\ee
Grouping the four strands of a line defines an operator,  $\TT$,  that acts separately and
independently on each strand of the propagator:
\be  \TT = \oplus_{j_{f}}    \otimes_{f=1}^4
 T_{j_f}^{\gamma}
\ee
so that the EPRL/FK propagator, to be compared with \eqn{fancyprop},  is
\be
C= \jj \TT \jj \label{defC}
\ee
that is
\beqa
C(g,g')&=&  \int du dv
\prod_{f=1}^4\sum_{j_f} \bigl[  \delta_{j_{f-}/j_{f+} = (1-\gamma) /(1+ \gamma) }  \bigr]
\alpha_{j_f} \beta_{j_f}\int dn_f \\
&&{\mathrm{Tr}}_{j_{f+}\otimes j_{f-}} \left( u g_f \;
(g^{'}_f)^{-1}v^{-1} |j_{f+},n_f>\otimes |j_{f-},n_f\rangle \langle
j_{f+},n_f|\otimes<j_{f-},n_f| \right),\nonumber
\eeqa
with  $ \alpha_{j}=d_{j_+}d_{j_-},\
\beta_j = d_{j_{+} + j_{-}}.
$
We state, without  proof, a number of properties of the propagator, for more details
see \cite{noi, GGR}
\begin{itemize}
\item The operator $C$  is hermitian
\item $\TT$ is a projector
\item Since the propagator is hermitian, Feynman amplitudes are
independent of the orientations of faces and propagators.
\item Since $\TT$ and $\jj$ do not commute, the propagator $C$ can
have non-trivial {\it spectrum} (with eigenvalues between 0 and 1).
\item Slicing the eigenvalues should allow a renormalization group
analysis. This is why we  call this kind of theories {\it dynamic} GFT's.
\item Since $\TT$ is a projector, the propagator $C$ of the EPRL/FK
theory is bounded in norm by the propagator of the $BF$ theory, as well as Feynman
amplitudes.
\end{itemize}
\subsubsection{EPRL/FK Amplitudes}
Combining the propagator and the vertex expressions, the integrations over all $g,g'$
group variables can be performed explicitly, leading to the amplitude of any  graph
${\mathcal G}$. This amplitude is  given by an integral of a product over all faces of
the graph as in \eqref{amp8}, but the amplitudes for faces are different. To compute
these  face amplitudes we distinguish between closed faces (no external edges) and open
faces (which end on external edges). For example, the self-energy graph in fig.
\ref{2P-fig} has 6 closed faces and 4 open faces. Using the same numbering of the  $p$
edges and vertices along a closed face, its amplitude is given by
\beqa
\mathcal{A}_f=\int \prod_{a=1}^p \bigl(dg_{\ell_a} dg'_{\ell_a}
\bigr) \sum_{j_{\ell_a}} \alpha_{j_{\ell_a}}
 \tr_{j_{\ell_a} +\otimes j_{\ell_a} -} \bigl( (u_{\ell_a}  g_{\ell_a}
(g'_{\ell_a})^{-1}v_{\ell_a}^{-1})^{\eta_{\ell_a f}}
T_{j_{\ell_a}}^\gamma \bigr) \prod_v V_v,
\label{faceampl0}
\eeqa
where the constraint on $j_+, j_-$ is implicitly understood from now
on and $V_v$ are the vertices.
 To perform the integrals over the strand variables $g, g'$ we use the standard results
 \beqa
\int d g \, {{\overline{R}^{(j)}}_n} \,^m (g) {R^{(j') p}}_q
(g)&=&
\frac{1}{d_j}\delta(j,j') \delta^m_q \delta_n^p \label{intgg}\\
\int d g \, {R^{(j) m}}_n (g) {R^{(j')p}}_q (g)&=&
\frac{1}{d_j}\delta(j,j') \epsilon^{m p}
\epsilon_{nq}\label{intgg-1}
\eeqa
with ${R^{(j)m}}_n(g)$  the matrix element of $g$ in the $2j+1$ dimensional  representation and     ${{\overline{R}^{(j)}}_n} \,^m (g)={R^{(j)m}}_n(g^{-1})$ which
imply
\beqa
\int d g \tr_j Ag\,\tr_{j'}
Bg^{-1}&=&\frac{1}{d_j}\delta(j,j')\tr_j
AB
\label{gg-1}\\
\int d g \tr_j Ag\,\tr_{j'} Bg&=&\frac{1}{d_j}\delta(j,j')\tr_j
A\epsilon B^T\epsilon^T
\label{gg}
\eeqa
and  $\epsilon\in SU(2)$,
\be \label{epsilonop}
\epsilon= \left( \begin{array}{cc}
0 & 1 \\
 -1 & 0 \\
\end{array}
 \right).
\ee
Thus we arrive at
\be
\mathcal{A}_f =\sum_{j_f} \alpha_{j_f} \tr_{j_{f+}\otimes j_{f-}}
\overrightarrow{\prod}_{\stackrel{a=1}{}}^p \left(h_{\ell_a,v_a}
^{\eta_{\ell_a f}}
 h_{\ell_{a+1}, v_{a}}^{\eta_{\ell_{a+1}f}}
T_{j_f}^\gamma\right),
 \label{faceampl}
\ee
with $\ell_{p+1}=\ell_1$. It is interesting to notice that we recover the $SU(2)$ BF
theory in the limit $\gamma \rightarrow 1$ (see \cite{noi} for details). For open faces
(which end on external edges) we obtain  a slightly modified expression  but we also find
that their contribution is irrelevant in the large spin approximation. It can be shown
that the graph amplitude obtained by replacing \eqn{faceampl} in \eqn{amp8} is the same as the spin foam amplitude found in \cite{ELPR} when  passing to spin representation.
\section{The stationary phase method}
In \cite{noi} we have evaluated  the degree of divergence of the graphs for GFT models
using the stationary phase method. Here we give a brief account of the calculation, which is however quite technical, and we state the results. We refer to the original paper for more details.
To this analysis, we have
 to rewrite the graph amplitude
\eqn{amp8}, with face amplitudes depending on the model, in the form
\be
{\cal A}_{\cG}=\sum_{j_{f} \le \Lambda}{\cal N}\int \prod dh \prod
dn\exp\big\{\sum_{f}j_{f}S_{f}[h,n]\big\},
\ee
with
 ${\cal N}={\cal N}(j) $ and   $\Lambda$ some ultraspin cutoff.
To evaluate the superficial power
counting, we set $j_{f}=jk_{f}$ with $k_{f}\in[0,1]$ and use the stationary phase method
to derive the large $j$ behavior of the summands. This calculation has been performed
both for BF models and for the EPRL/FK model.
 If the action is complex but has a negative real part, the
contribution to this integral are quadratic fluctuations around
zeroes of the real part of $S$ which are stationary points of its
imaginary part, otherwise the integral is exponentially suppressed
as $j\rightarrow\infty$.

\subsection{ {The saddle point method for 3D BF models}}
For 3D BF models the degree of divergence of the graphs has already been calculated in the literature \cite{FGO, GKMR}. We re-derive such results  to test our method.
To write the action $S_f$ we introduce the projector
\be
|n\rangle\langle n|={\textstyle \frac{1}{2}}\big(1+\sigma\cdot \label{cohproj}
n\big),
\ee
with $|n\rangle\equiv|n,\frac{1}{2}\rangle$,
and observe that
 \be\langle n,j|g|n',j\rangle=\langle n|g|n'\rangle^{2j} \label{gnn'}
\ee
so that
\be
S_{f}[h,n]=k_{f}\log\mbox{Tr}\Big[\big(\mathop{\prod}\limits_{\ell\in\partial
f}^{\longrightarrow}h_{\ell}^{\eta_{\ell,f}}\big)\big(1+\sigma\cdot
n\big) \Big].
\ee
 Since the action is the logarithm of the trace of the product
of a unitary element and a projector,  its real part is negative and
maximal  at $h_{\ell}=1$.
 To perform the saddle point expansion, we
expand the group element to second order as
\be
h_{\ell}=1-\frac{A_{\ell}^{2}}{2}+\mathrm{i}\,A_{\ell}\cdot\sigma+O(A_{\ell}^{3})
\ee
with $A\in su(2)$. Also
\be
n_{f}=n_{f}^{(0)}+\xi_{f}-\frac{\xi_{f}^{2}}{2}n^{(0)}_{f}+O(\xi_{f}^{3}),\qquad\mbox{with}\quad
n_{f}^{(0)}\cdot\xi_{f}=0.
\ee
(because $n_{f}^{2}=1$ up to third order terms).
 Let us
consider a face with edges $\ell_{1},\dots,\ell_{p}$, then to second
order
\be
\mathop{\prod}\limits_{\ell\in\partial
f}^{\longrightarrow}h_{\ell}^{\eta_{\ell,f}}
=1-\frac{A_{f}^{2}}{2}+\mathrm{i}\,\sigma\cdot
A_{f}-\mathrm{i}\,\sigma\cdot\Phi_{f}
\ee
with
\be
A_{f}=\sum_{1\leq a\leq
p}\eta_{\ell_{a},f}A_{\ell_{a}}\qquad\mbox{and}\qquad
\Phi_{f}=\sum_{1\leq a<b\leq p}\eta_{\ell_{a},f}\eta_{\ell_{b},f} \,
A_{\ell_{a}}\wedge A_{\ell_{b}}.
\ee
this implies
\be
S_{f}[A_{\ell},\xi_{f}]=2k_{f}\Big\{\mathrm{i}\,n_{f}\cdot
A_{f}-\frac{A_{f}^{2}}{2}+\frac{(n_{f}\cdot
A_{f})^{2}}{2}+\mathrm{i}\,\xi_{f}\cdot
A_{f}+\mathrm{i}\,n_{f}\cdot\Phi_{f}\Big\}
\ee
 and we have to estimate
\be
\int\prod_{\ell}dA_{\ell}\prod_{f}d\xi_{f}\,\exp 2j \sum_{f}
k_{f}\Big\{\mathrm{i}\,n_{f}\cdot
A_{f}-\frac{A_{f}^{2}}{2}+\frac{(n_{f}\cdot
A_{f})^{2}}{2}+\mathrm{i}\,\xi_{f}\cdot
A_{f}+\mathrm{i}\,n_{f}\cdot\Phi_{f}\Big\}
\ee
as $j\rightarrow\infty$.

We do not integrate over the vectors $n_{f}$. They  have
to be chosen so that they are extrema of the imaginary part of $S$.

Because all terms except the first one
$\sum_{f}k_{f}n_{f}\cdot A_{f}$ are of second order, the imaginary
part is stationary if and only if
\be
\sum_{f} \mathrm{i}\,k_{f}n_{f}\cdot A_{f}=\sum_{\ell,f}
\mathrm{i}\,\eta_{\ell,f}k_{f}n_{f}\cdot A_{\ell}=0 \qquad\forall
A_{\ell}\in{\Bbb R}^{3},
\ee
which amounts to the closure condition
\be
\sum_{f}\eta_{\ell,f}k_{f}n_{f}=0, \; \; \;\forall \ell \label{clo}
\ee
  This is the
requirement that, in the semi-classical limit,  the vectors
$j_{f}n_{f}$ are the sides of a triangle (resp. the area bivectors
of a tetrahedron) that propagates along $\ell$ in $D=3$ (resp.
$D=4$).

The solutions of the closure condition \eqn{clo} range from non degenerate to
maximally degenerate.
In three dimensional (resp. four dimensional ) BF theory,  a
solution is said to be {\it non degenerate} if all the
tetrahedra (resp. 4-simplices) corresponding to the vertices of the
graph have maximal dimension.
 At the opposite end, we say
that a solution is {\it maximally degenerate} if all the vectors
$n_{f}$ are proportional to a single one $n_{0}$,
\be
n_{f}=\sigma_{f}n_{0}\qquad\mbox{with}\qquad
\sigma_{f}\in\left\{-1,+1\right\}.
\ee
In both cases we find
\be
{\cal A}_{\cG}\sim \Lambda^{3F-3r}
\ee
with $r$ the rank of the $L\times F$ incidence matrix
$\eta_{\ell,f}$.

In particular for the self-energy graph in fig. (\ref{2P-fig}) we find
\be
{\cal A}_{\cG_2} \sim \Lambda^9
\ee
with $F=6$, the number of closed faces. We do not include open faces because they do not contribute to the divergence of the graph, as we will argue below.
\subsection{The self-energy in the EPRL/FK
model}
Let us come to the EPRL/FK model. An analysis for generic graphs is possible although technically difficult.  Here we only consider   the self-energy graph $\cG_2$ of fig. (\ref{2P-fig}), which represents the first quantum correction to the propagator. Moreover, we limit the analysis to non-degenerate configurations, as in \cite{PRS}. For completely degenerate configurations the stationary phase method only provides upper bounds. The issue is discussed in \cite{noi}.

The  self-energy graph $\cG_2$ has $4$ open faces and $6$ closed faces with two edges. We label the
internal propagators with an index $a$ ranging from  $1$ to  $4$ and
orient them in the same direction. We label the $6$ closed faces
with pairs of indices $(a,b)$, $a<b$. We do not worry about  open faces in our analysis, which technically amounts to put to zero the external spins.  However, the result should also hold  for finite non-zero spins on the external faces. Indeed, since the latter remain finite, the contribution of the external faces to the action can be neglected
as $j\rightarrow\infty$.
The amplitude of the self-energy graph reads then
\beqa
{\cal A}_{\cG_2}&=&\prod_a d u_a^{\pm} {d v_a}^{\pm}\prod_{a<b}
\mathcal{A}_{ab}
\eeqa
where, from \eqn{faceampl} the face amplitude reads
\beqa
\mathcal{A}_{ab}&=&\sum_jd_{j_+} d_{j_-} \beta_j^2 \int d n_{ab} d
n_{ab}'  \langle j_+ n_{ab}|u_{a +} u_{b +}^{-1}|j_+ n_{ab}'\rangle
\langle j_+ n_{ab}'|v_{b +} v_{a +}^{-1}|j_+ n_{ab}\rangle
\nonumber\\
&\times&\;\;\;\;\langle j_- n_{ab}|u_{a -} u_{b -}^{-1}|j_-
n_{ab}'\rangle\langle j_- n_{ab}'|v_{b -} v_{a -}^{-1}|j_-
n_{ab}\rangle.
\eeqa
In order to perform a stationary phase analysis we rewrite the graph
amplitude as
\be
{\cal A}_{\cG_2}=\sum_{j_{f}}\int
\prod_{a}du_{a}^{\pm}\,\prod_{a}dv_{a}^{\pm}\,\prod_{i}dn_{i}\,
\prod_{f}\Big\{(d_{j_{f}})^{2}d_{j^{+}_{f}}d_{j^{-}_{f}}\exp\big\{jS^{+}_{f}+jS^{-}_{f}\big\}\Big\}\label{self-energy}
\ee
with $j_{f}^{\pm}=j\gamma^{\pm}k_{f}$, $k_{f}\in[0,1]$ and $j$
large. The face action for $f=ab$ is then
\be
S^{\pm}_{f}= 2\gamma^{\pm}k_{f}\log\big\{\langle
n_{f,a}|u_{a}^{\pm}(u_{b}^{\pm})^{-1}|n_{f,b}\rangle\langle
n_{f,b}|v_{b}^{\pm}(v_{a}^{\pm})^{-1}|n_{f,a}\rangle\big\}
\ee
 We employ the saddle point technique around the identity
\be
u_{a}^{\pm}=1-\frac{(A_{a}^{\pm})^{2}}{2}+\mathrm{i}\,\sigma\cdot
A_{a}^{\pm}+O(A_{a}^{\pm})^{3},\;\;
v_{a}^{\pm}=1-\frac{(B_{a}^{\pm})^{2}}{2}+\mathrm{i}\,\sigma\cdot
B_{a}^{\pm}+O(B_{a}^{\pm})^{3}
\ee
and use the result \eqn{cohproj} so that the action at the identity for the face $f=ab$
reads
\be
S^{\pm}_{f}[1,1,n_{i}]=
\gamma^{\pm}k_{ab}\log\Big\{\frac{1+n_{f,a}\cdot n_{f,b}}{2}\Big\}
\ee
which is negative except for $n_{f,a}=n_{f,b}=n_{f}$.

We perform the expansion of the coherent state
around a unit vector common to all the strands of the face
\be
n_{i}=n_{f}+\xi_{i}-\frac{(\xi_{i})^{2}}{2}n_{f}+O(\xi_{i})^{3},\qquad
\mbox{with}\qquad n_{f}\cdot\xi_{i}=0,
\ee
otherwise the integral is exponentially damped.
It is convenient to perform the following change of
variables
\be
A^{\pm}_{a}=A_{a}\pm \gamma^{\mp}X_{a}\qquad\mbox{and}\qquad
B^{\pm}_{a}=B_{a}\pm \gamma^{\mp}Y_{a}
\ee
Terms linear in $A^{\pm}$ and $B^{\pm}$  only involve $A$
and $B$, while in the quadratic terms, the pair of variables $A$ and
$B$ on one side and the pair $X$ and $Y$ on the other side decouple.

Thus we observe that,
at the level of the quadratic approximation, we can  separate the action into a SU(2) BF action
(variables $A$ and $B$) and an ultralocal potential that only
involves uncoupled variables attached to the vertices (variables $X$
and $Y$). Moreover it can be shown that this result is true for a generic graph.

 We  perform the
Gaussian integration over the two dimensional vector
$\chi_{f}=\xi_{f,a}-\xi_{f,b}$, and obtain
\beqa
{\cal A}_{\cG_2}&=&\sum_{j_{f}}\,j^{18}\,\Big\{ \int
\prod_{a}dA_{a}\,\prod_{a}dB_{a}\,\prod_{f}d\xi_{f}\exp
jS_{BF}(A,B,\xi)\\ &&\times\int \prod_{a} dX_{a}\exp jQ(X)\times\int
\prod_{a} dY_{a}\exp jQ(Y)\Big\} \eeqa with
$\xi_{f}=\xi_{f,a}+\xi_{f,b}$.
The  BF-like action is
\beqa
S_{BF}[A,B,\xi]&=&\sum_{a<b}k_{ab}\Big\{-\frac{1}{2}\Big[n_{f}\wedge\big(A_{a}-A_{b}+B_{b}-B_{a}\big)\Big]^{2}\\
&&+\mathrm{i}\,n_{ab}\cdot\big(A_{a}-A_{b}+B_{b}-B_{a}\big)+\mathrm{i}\,n_{ab}\cdot\big(A_{a}\wedge
A_{b}+B_{b}\wedge B_{a}\big)\\
&&+\mathrm{i}\,\xi_{ab}\cdot\big(A_{a}-A_{b}+B_{b}-B_{a}\big) \Big\}
\eeqa
while the ultra local terms are
\be
Q[X]=\gamma^{+}\gamma^{-}\sum_{a<b}k_{ab}\Big\{\big[n_{ab}\wedge(X_{a}-X_{b})\big]^{2}+\mathrm{i}\,n_{ab}\cdot\big(X_{a}\wedge
X_{b}\big) \Big\} . \label{QX}
\ee
The Gaussian integral over the variables $A$ and $B$ can be
evaluated using the same techniques as for the pure BF models, thus yielding
\begin{equation}
\int \prod_{a}dA_{a}\,\prod_{a}dB_{a}\,\prod_{f}d\xi_{f}\exp jS_{BF}(A,B,\xi)\quad\sim\quad j^{-9}\label{ABself}
\end{equation}
as $j\rightarrow \infty$.
As for the  Gaussian integrals over the independent variables
$X_{a}$ and $Y_{a}$, we have
\be
\int \prod_{a} dX_{a}\exp jQ(X)\quad\sim\quad j^{-\frac{\mbox{\tiny
rank}(Q)}{2}}
\ee
with $Q[X]$ the quadratic form in Eq. \eqn{QX} and a similar expression for $Q[Y]$.
The rank  may be computed to be rank(Q)= 9 so that the Gaussian integral over $X, Y$ yields
a power of $j^{-9/2}$ each.
Therefore, we obtain the power counting for the self-energy with non
degenerate configurations as follows
\be
\sum_{\mbox{\tiny 6 independent spins }\sim j\,<\Lambda}\,
j^{24}\times j^{-6}\times j^{-9}\times \big(j^{-9/2}\big)^{2} \sim
\Lambda^{6},
\ee
where
\begin{itemize}
\item $j^{24}$ arises from a $d_{j^{+}}d_{j^{-}}\sim
j^{2}$ for each of the 6 faces
\item   $d_{j}\sim j$ for
each of the two strands in each face
\item  $j^{-6}$
results from the Gaussian integration over the 6 variables
$\chi_{f}=(\xi_{f,a}-\xi_{f,b})$
\item  $j^{-9}$ from the
integration over $A$ and $B$
\end{itemize}
 This reproduces the result of [PRS], with non degenerate configurations.

\end{document}